\documentclass[aps,prc,twocolumn,superscriptaddress,showpacs,floatfix,preprintnumbers]{revtex4}
\usepackage{graphicx,color}
\usepackage{braket}
\usepackage{amsmath}
\usepackage{amssymb}
\usepackage{multirow}
\usepackage{ulem}

\newcommand{\Ve}[1]{\ensuremath{\boldsymbol{#1}}}
\newcommand{\be}{\begin{equation}}
\newcommand{\ee}{\end{equation}}
\newcommand{\eq}[1]{\begin{align}#1\end{align}}
\begin{document}
\preprint{NITEP 78}
\title{Practical method for decomposing discretized breakup cross sections\\
into components of each channel}

\author{Shin Watanabe}
\email[]{s-watanabe@gifu-nct.ac.jp}
\affiliation{National Institute of Technology, Gifu College, Gifu 501-0495, Japan}
\affiliation{RIKEN, Nishina Center, Wako, Saitama 351-0198, Japan}

\author{Kazuyuki Ogata}
%\email[]{kazuyuki@rcnp.osaka-u.ac.jp}
\affiliation{Research Center for Nuclear Physics (RCNP), Osaka University, Ibaraki 567-0047, Japan}
\affiliation{Department of Physics, Graduate School of Science, Osaka City University, Osaka 558-8585, Japan}
\affiliation{Nambu Yoichiro Institute of Theoretical and Experimental Physics (NITEP), Osaka City University, Osaka 558-8585, Japan}

\author{Takuma Matsumoto}
%\email[]{matsumoto@phys.kyushu-u.ac.jp}
\affiliation{Department of Physics, Kyushu University, Fukuoka 812-8581, Japan}

\begin{abstract}
\noindent{\bf Background:}
In the continuum-discretized coupled-channel method, a breakup cross section (BUX) is
obtained as an admixture of several components of different channels in multichannel scattering.

\noindent{\bf Purpose:}
Our goal is to propose an approximate way of decomposing the discretized BUX into components of each channel.
This approximation is referred to as the ``probability separation (P separation)."

\noindent{\bf Method:}
As an example, we consider $^{11}$Be scattering by using the three-body model with core excitation ($^{10}\mathrm{Be}+n+T$, where T is a target).
The structural part is constructed by the particle-rotor model and the reaction part is described by the distorted-wave Born approximation (DWBA).

\noindent{\bf Results:}
The validity of the P separation is tested by comparing with the exact calculation.
The approximate way reproduces the exact BUXs well regardless of the configurations and/or the resonance positions of $^{11}$Be.

\noindent{\bf Conclusion:}
The method proposed here can be an alternative approach for decomposing discretized BUXs into components
 in four- or five-body scattering where the strict decomposition is hard to perform.

\end{abstract}

\date{\today}

\maketitle

{\it Introduction.}
A breakup cross section (BUX) is an important observable
 to investigate not only nuclear structures, but also reaction dynamics.
Theoretically, BUXs have been calculated by using various reaction models,
such as the adiabatic approximation~\cite{Joh70},
the Glauber model~\cite{Glauber},
the semiclassical model~\cite{Ber92,Kid94},
the continuum-discretized coupled-channel method (CDCC)~\cite{CDCC-review1,CDCC-review2,CDCC-review3},
and the Faddeev formalism~\cite{Del07}.
CDCC is one of the most powerful and flexible methods of describing
breakup processes induced by weakly bound nuclei.
In the 1980s, three-body CDCC was first applied to
 $d$-induced reactions on a target nucleus ($T$), where the $n+p+T$ three-body model was assumed.
Three-body CDCC has been successful in describing many kinds of three-body reactions.
Nowadays, three-body CDCC has been developed mainly in two directions.
One is four-body CDCC~\cite{Mat04,Mat06,Mat10,4body-CDCC-bin,HRWF,Des18}
and the other is three-body CDCC with core excitation~\cite{Die14,Lay16,Die17}.
These methods address breakup reactions including multi-BU channels.
For example, for $^6$Li scattering ($n+p+\alpha+T$),
four-body CDCC should take into account the three- and four-body channels,
\eq{
{}^6\mathrm{Li}+T&\rightarrow d+\alpha+T\hspace{2mm}\mathrm{(three\mathchar`-body\ channel)},\label{eq:3b-channel}\\
{}^6\mathrm{Li}+T&\rightarrow n+p+\alpha+T\hspace{2mm}\mathrm{(four\mathchar`-body\ channel)}.\label{eq:4b-channel}
}
These channels are coupled to each other during the scattering and should be treated on an equal footing.
Furthermore, each of the BUXs should be separately calculated in the multi-BU channel.

In four-body CDCC including the multi-BU channels of $^6$Li scattering, 
the pseudostate method~\cite{Mat03,Mat04,Wat15,Des15,Cas15}
is a reasonable way of calculating discretized-continuum states.
In the pseudostate method, the projectile wave functions are constructed by diagonalizing
the internal Hamiltonian of the projectile with the $L^2$-basis functions.
Hence, it is not necessary to solve the scattering problem of
the $n+p+\alpha$ three-body system under the proper three-body boundary conditions.
The pseudostates thus obtained are energetically discretized and
consist of both the $d\alpha$-component (three-body channel) and the $np\alpha$-component (four-body channel) with a certain weight.
As a result, CDCC with the pseudostate method describes the transition between these mixed channels.
Furthermore, in Ref.~\cite{Mat10}, a new method has been proposed to construct a continuous 
breakup cross section regarding the breakup energy, which enables one to directly compare the 
result of four-body CDCC with experimental data. However, it is not possible for the pseudostate 
method to disentangle the cross section into three-body and four-body channel components.

Recently, a new method utilizing the solution to the complex-scaled Lippmann-Schwinger equation (CSLS)
was proposed and applied to the $d(\alpha,\gamma){}^6$Li radiative capture process~\cite{Kik11}.
 A remarkable feature of this method is the specification of the incident channel in solving the 
three-body scattering problem in the space defined by a complex-scaled Hamiltonian. Once it is implemented
 into a four-body CDCC calculation, one may obtain continuous breakup observables with a clear separation of
 the three-body and four-body channel components. However, such a CDCC-CSLS calculation is rather numerically
 demanding and has been limited to $^6$He scattering ($n+n+\alpha+T$)~\cite{Kik13}
 to which only the four-body channels are relevant.
There are also several four-body CDCC calculations for extracting information on three-body
 continuum states~\cite{Cas19,Des15,Cas15},
but they focus only on the scattering of Borromean systems, such as $^9$Be and $^{16}$Be,
and there is no need to decompose BUXs into each component.
In this Letter, we propose a practical method for decomposing the discretized BUXs into each BUX component.
This method is referred to as the ``probability separation (P separation)" from now on.
The P separation does not require the exact solutions or the smoothing procedure.
In the previous work~\cite{Wat15}, the P separation was applied to the analysis of
$^6$Li elastic scattering and successful in separating each of the three- and
four-body channel-coupling effects on the elastic cross sections.
Thus, it is expected that the P separation is applicable for separating the BUXs as well.
It is worth noting that the P separation was used for getting rid of the $d\alpha$- or $np\alpha$-dominant pseudostates
before reaction calculations in the previous work~\cite{Wat15},
whereas the P separation is applied for separating the discretized BUX after reaction calculations in the present Letter.

Before going to four-body scattering, we first consider $^{11}\mathrm{Be}+T$ three-body scattering with core excitation~\cite{Cre11}
because this scattering provides an analogy to the $^6$Li four-body scattering regarding the mixture of different channels.
In this case, we consider two BU channels as
\eq{
{}^{11}\mathrm{Be}+T&\rightarrow {}^{10}\mathrm{Be(g.s.)}+n+T\hspace{2mm}\mathrm{(core\mathchar`-ground\ channel)},\\
{}^{11}\mathrm{Be}+T&\rightarrow {}^{10}\mathrm{Be}^*+n+T\hspace{2mm}\mathrm{(core\mathchar`-excited\ channel)},
}
where g.s. represents the ground state.
As a merit of this scattering, we can easily obtain the exact breakup wave functions of each channel
for the $^{11}$Be two-body projectile unlike the $^6$Li three-body projectile.
In the actual analysis, the projectile wave functions are constructed by the particle-rotor model~\cite{Boh75,Ura11,Ura12},
and the BUXs are calculated by the extended distorted-wave Born approximation (xDWBA)~\cite{Mor12-PRC,Mor12-PRL}.
The xDWBA enables us to calculate both the exact (continuous) and the approximate (discretized) $T$-matrix
elements as shown later.
We can then compare the approximate BUXs with the exact ones quantitatively.
After validating the P separation, we apply it to $^6$Li scattering
and predict the $d\alpha$ and $np\alpha$ BUXs, respectively.

{\it Theoretical framework.}
The three-body Hamiltonian with core excitation is given by
\eq{
H&=K_{\Ve R}+V_{vT}(R_{vT})+V_{cT}({\Ve R}_{cT},{\Ve \xi})+h_{P},\label{eq:H}\\
h_{P}&=K_{\Ve r}+V_{vc}({\Ve r},{\Ve \xi})+h_{c}({\Ve \xi}),\label{eq:h_p}
}
where $\Ve R$ is the relative coordinate between the center of mass of a projectile ($P$) and a target ($T$),
${\Ve r}$ represents the relative coordinate between a valence neutron ($v$) and a core nucleus ($c$),
and ${\Ve \xi}$ stands for the internal coordinate of $c$, i.e., the angle of the symmetric axis of the deformed core.
The operators $K_{\Ve R}$ and $K_{\Ve r}$ are the kinetic energies associated
with $\Ve R$ and $\Ve r$, respectively, $V_{ab}$ (${a,b}={T,v,c}$) is the interaction between a and b,
and $h_{c}$ is the internal Hamiltonian of the core.
Here, $V_{cT}$ and $V_{vc}$ are assumed to be nonspherical,
which can trigger the core excitation.

The projectile wave function is constructed with the particle-rotor model by solving the Schr\"{o}dinger equation,
\be
h_{P}\Psi^{(\alpha)}_{JM}({\Ve r},{\Ve \xi})=\varepsilon\Psi^{(\alpha)}_{JM}({\Ve r},{\Ve \xi}),\label{eq:Sch-proj}
\ee
where $\Psi^{(\alpha)}_{JM}$ is the projectile wave function with the total angular momentum $J$ and its projection $M$,
$\varepsilon$ is the eigenenergy,
and $\alpha$ represents the initial $(i)$ or the final ($f$) channel defined later.
$\Psi^{(\alpha)}_{JM}$ is expanded as
\be
\Psi^{(\alpha)}_{JM}({\Ve r},{\Ve \xi})=\sum_{\ell jI}R^{(\alpha)}_{\ell jI}(r)[\mathcal{Y}_{\ell j}(\hat{\Ve r})\otimes \Phi_I({\Ve \xi})]_{JM}.
\ee
$\Phi_I$ represents the core state with the spin $I$, which
 satisfies the Schr\"{o}dinger equation $h_{c}\Phi_I=\epsilon_I\Phi_I$, where
$\epsilon_I$ is the eigenenergy of the core.
The coefficient $R^{(\alpha)}_{\ell jI}$ describes the relative motion between the core and the valence neutron,
where $\ell$ is the orbital angular momentum and $j$ is the total angular momentum.

The breakup $T$-matrix elements in xDWBA are given by
\eq{
T_{fi}^{JM,J_0M_0}
&=\Bra{\chi_{{\Ve K}}^{(-)}({\Ve R})\Psi_{JM}^{(f)}({\Ve r},{\Ve \xi})}V_{vT}(R_{vT})\nonumber\\
&\hspace{10mm}+V_{cT}({\Ve R}_{cT},{\Ve \xi})
\Ket{\chi_{{\Ve K}_0}^{(+)}({\Ve R})\Psi_{J_0M_0}^{(i)}({\Ve r},{\Ve \xi})},\label{eq:Tfi}
}where
$\chi_{{\Ve K}_0}^{(+)}$ ($\chi_{{\Ve K}}^{(-)}$) is the initial outgoing (final incoming) distorted wave
with the initial (final) wave number ${\Ve K}_0$ (${\Ve K}$),
$\Psi_{J_0M_0}^{(i)}$ and $\Psi_{JM}^{(f)}$ are the initial $(i)$ and the final ($f$) projectile
wave functions, respectively.
The initial wave function $\Psi_{J_0M_0}^{(i)}$ is nothing but the ground state wave function,
whereas the final wave function $\Psi_{JM}^{(f)}$ represents a continuum state with the incoming asymptotic form.
As for $\Psi_{JM}^{(f)}$, the neutron and core states in the asymptotic region
 should be specified by not only the eigenenergy, but also the angular momentum, the total angular momentum,
 and the spin of the core, i.e., $f=\{\varepsilon_f,\ell_f,j_f,I_f$\}.
If the excitation energy $\varepsilon_f$ is not large enough,
core-excited channels $f=\{\varepsilon_f,\ell_f,j_f,I_f\ne0$\} are closed.
This will be discussed at around Eq.~\eqref{eq:imposition}.

For the explanation below, the final channel
$f=\{\varepsilon_f,\ell_f,j_f,I_f$\} is explicitly shown in the $T$-matrix element,
\be
T_{fi}^{JM,J_0M_0}\rightarrow T_{\{\ell_f,j_f,I_f\},i}^{JM,J_0M_0}(\varepsilon_f)\hspace{5mm}\mathrm{(exact)}.\label{eq:exact-T}
\ee
If Eq.~\eqref{eq:Sch-proj} is solved by the diagonalization method, 
not only the different energies $\varepsilon_f$, but also the different sets of $\{\ell_f,j_f,I_f\}$ are superposed
as a pseudostate $\hat{\Psi}_{JM}^{(n_f)}$,
where ``$\hat{\hspace{3mm}}$" denotes discretization and $n_f$ is the state number. 
By replacing $\Psi_{JM}^{(f)}$ with $\hat{\Psi}_{JM}^{(n_f)}$ in Eq.~\eqref{eq:Tfi},
the exact $T$-matrix element is discretized as
\be
T_{fi}^{JM,J_0M_0}\rightarrow \hat{T}_{n_fi}^{JM,J_0M_0}\hspace{5mm}\mathrm{(discretized)}\label{eq:discretized-T}.
\ee
Note that the discretization is not indispensable for the DWBA but it is for CDCC.

From Eqs.~\eqref{eq:exact-T} and \eqref{eq:discretized-T},
the exact and the discretized BUXs are given by
\eq{
\frac{d^2\sigma_{\{\ell_f,j_f,I_f\}}(\varepsilon_f)}{d\Omega d\varepsilon_f}&=\gamma\sum_{MM_0}\left|T_{\{\ell_f,j_f,I_f\},i}^{JM,J_0M_0}(\varepsilon_f)\right|^2,\\
\frac{d\hat{\sigma}_{n_f}}{d\Omega}&=\gamma\sum_{MM_0}\left|\hat{T}_{n_fi}^{JM,J_0M_0}\right|^2,
}
respectively, where $\Omega$ is the solid angle of ${\Ve K}$ and $\gamma$ is the kinematic factor.
 It is worth noting that
 $d\hat{\sigma}_{n_f}/d\Omega$ is specified only by the final-state number $n_f$.
In the following analysis, we compare the energy-integrated total BUXs,
\be
\sigma(\mathrm{tot})=\sum_{I_f}\sigma(I_f),
\ee
with
\be
\sigma(I_f)=\sum_{\ell_f,j_f}\int d\Omega d\varepsilon_f 
\frac{d^2\sigma_{\{\ell_f,j_f,I_f\}}(\varepsilon_f)}{d\Omega d\varepsilon_f},
\ee
and the energy-summed total BUXs,
\eq{
\hat\sigma(\mathrm{tot})&=\sum_{n_f}\int d\Omega \frac{d\hat{\sigma}_{n_f}}{d\Omega}.
}
Henceforth, we refer to $\sigma(I_f=0)$ and $\sigma(I_f\ne0)$ as
the ``core-ground BUX" and the ``core-excited BUX," respectively.

In order to decompose the discretized total BUX $\hat\sigma(\mathrm{tot})$ into
 the approximate core-ground and core-excited BUXs with the P separation,
we first derive the core-ground and core-excited probabilities
by taking the overlap between $\hat{\Psi}_{JM}^{(n_f)}$ and $[\mathcal{Y}_{\ell j}\otimes \Phi_{I}]_{JM}$ as
\be
P_{n_f}(I)
=\int_0^\infty r^2dr\sum_{\ell j}\left|\Braket{[\mathcal{Y}_{\ell j}\otimes \Phi_{I}]_{JM}|\hat{\Psi}_{JM}^{(n_f)}}\right|^2.\label{eq:prob}
\ee
We can then define each of the approximate total BUXs by
\be
\hat{\sigma}(I_f)=\sum_{n_f} \hat{\sigma}_{n_f}(I_f)\hspace{2mm}\mathrm{with}\hspace{2mm}
\hat{\sigma}_{n_f}(I_f)=P_{n_f}(I_f) \hat{\sigma}_{n_f},\label{eq:approx1}
\ee
which satisfies the relation 
\be
\hat{\sigma}(\mathrm{tot})=\sum_{I_f}\hat{\sigma}(I_f).\label{eq:relation}
\ee
However, in this approximation, the core-excited BUX can be finite even below
 the ${}^{10}\mathrm{Be}(2_1^+)+n$ breakup threshold energy $(\varepsilon_\mathrm{th}=3.368$ MeV)
because of the finite core-excited components.
Therefore, we imposed the relations
$\hat{\sigma}_{n_f}(0)=\hat{\sigma}_{n_f}$ and $\hat{\sigma}_{n_f}(I_f\ne0)=0$
below the threshold energy. Namely, the condition,
\be
P_{n_f}(0)=1\ \mathrm{and}\ P_{n_f}(I\ne0)=0
\hspace{2mm} \mathrm{for}\hspace{2mm} \varepsilon_{n_f}<\varepsilon_\mathrm{th}\label{eq:imposition}
\ee
is added on Eq.~\eqref{eq:approx1}.
Equation~\eqref{eq:relation} is still satisfied under this condition.

In the actual calculation, we adopt the same potentials and the same model space used
 in Ref.~\cite{Mor12-PRC}. As a reaction part, 
the Gaussian interaction (depth $-45$ MeV and range 1.484 fm) is adopted for $V_{vT}$,
and the Watson potential~\cite{Wat69} is taken for the central part of $V_{cT}$.
As a structural part, the parity-dependent Woods-Saxon potential is taken;
radius $R_0=2.483$ fm, diffuseness $a=0.65$ fm, the central potential depth
 $V_0=-54.45$~MeV ($V_0=-49.61$ MeV) for the even (odd) parity
and the spin-orbit potential depth $V_\mathrm{so}=8.5$ MeV.
As for the $^{10}$Be core,  the deformation parameter $\beta_2=0.67$ is assumed,
and only two states are taken into account, i.e., the ground state ($I=0$)
and the first excited state ($I=2$, $\epsilon_2=3.368$ MeV).
The combination of this parameter set and the model space
well reproduces the energies
of the ground state ($J^\pi=1/2^+$, $\varepsilon_\mathrm{g.s.}=-0.50$ MeV),
the first excited state ($J^\pi=1/2^-$, $\varepsilon_\mathrm{first}=-0.18$ MeV),
and the several low-lying resonances~\cite{Mor12-PRC}.
The spin-parities $J^\pi=1/2^+$, $3/2^+$, $5/2^+$, $1/2^-$, and $3/2^-$ are considered.
In Eq.~\eqref{eq:Tfi}, the final distorted waves $\chi_{{\Ve K}}^{(-)}$ are evaluated by using the same potential
used for calculating $\chi_{{\Ve K}_0}^{(+)}$.

\begin{figure}[htbp]
\begin{center}
\includegraphics[width=0.45\textwidth,clip]{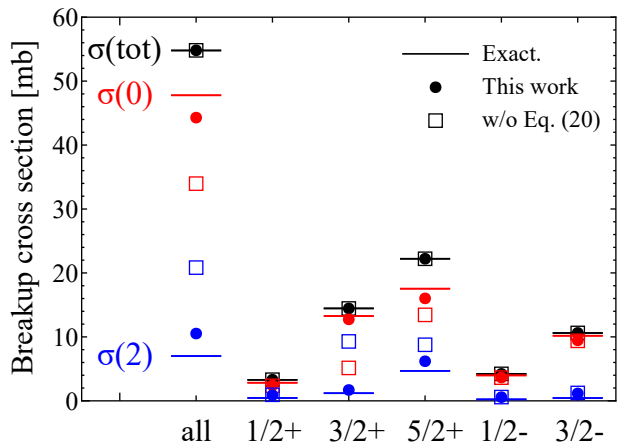}
\caption{Decomposition of the breakup cross sections for the scattering of $^{11}\mathrm{Be}+p$ at 63.7 MeV/nucleon.
The solid lines represent the exact BUXs ($\sigma$),
whereas the solid circles correspond to the approximate BUXs ($\hat{\sigma}$).
The results of the approximate decomposition without Eq.~\eqref{eq:imposition} are also
shown by the open squares.}
\label{fig:decomposition_both}
\end{center}
\end{figure}

{\it $^{11}$Be scattering with core excitation.}
First, the discretized total BUX $\hat\sigma(\mathrm{tot})$
is compared with the exact BUX $\sigma(\mathrm{tot})$
for the scattering of $^{11}\mathrm{Be}+p$ at 63.7 MeV/nucleon.
Both calculations give the same value of
$\hat{\sigma}(\mathrm{tot})=\sigma(\mathrm{tot})= 54.8$~mb, suggesting that the model space
is large enough for calculating the BUX.
Next, the discretized BUX is decomposed into the core-ground and the core-excited BUXs with the P separation.
The resultant BUXs are $\hat{\sigma}(0)=44.3$ and $\hat{\sigma}(2)=10.5$~mb,
whereas the exact BUXs are $\sigma(0)=47.8$ and $\sigma(2)=7.0$~mb.
Note that the smoothing calculation with
the discretized states gives the same result as the exact one.
In Fig.~\ref{fig:decomposition_both},
the approximate and the exact BUXs are further decomposed in terms of each spin parity of
$J^\pi=1/2^+$, $3/2^+$, $5/2^+$, $1/2^-$, and $3/2^-$.
The solid circles (solid lines) represent the approximate (exact) BUXs.
The total, core-ground, and core-excited BUXs are shown from the top to the bottom.
The approximate BUXs are in good agreement with the corresponding exact ones for each spin parity.
Thus, the P separation is found to work for the discretized BUX.

To confirm the validity of the P separation, we perform a systematic analysis.
We prepare other five sets of configurations
by changing the depth of potential and/or the excitation energy of ${}^{10}$Be
by following Ref.~\cite{Del16} (sets 1--6). The parameter sets are summarized in Table~\ref{tbl:summary}. 
Set 2 corresponds to the original parameter where the parity-dependent potential is taken.
For the other sets, the potential is common for all the states.
As summarized in Table~\ref{tbl:summary}, the total BUXs are decomposed into
$\hat{\sigma}(0)$ and $\hat{\sigma}(2)$ reasonably well regardless of the configurations.
For example, for set 6, small $\epsilon_2$ and large $P_\mathrm{gs}(2)$ may lead to the large core-excited BUX,
and the tendency is well reproduced by the P separation.
On the other hand, all the resonances constructed in sets 1--6 appear below the ${}^{10}\mathrm{Be}(2_1^+)+n$ threshold energy ($\varepsilon_\mathrm{th}$).
To construct the resonance(s) above $\varepsilon_\mathrm{th}$, we found that the deeper potential is necessary rather than the shallower potential.
If $V_0=-85.791$ and $\varepsilon_\mathrm{th}=0.5$ MeV are taken, two resonances are confirmed at around $1.1$ and $2.5$ MeV for the $5/2^+$ state.
However, at the same time, the original ground state becomes much deeper, i.e., the separation energy $S_n=15.25$ MeV.
We then use the newly made $1/2^+$ bound state with $S_n=1.0$ MeV as the ground state (set 7 in Table~\ref{tbl:summary}).
This is justified since we are comparing the P separation with the exact solution in theory.
As shown in set 7 of Table~\ref{tbl:summary}, the P separation works well even with the resonances above $\varepsilon_\mathrm{th}$.
The validity of the P separation is, thus, presented.

\begin{table*}[htbp]
\caption{Systematic analysis for the validity of the P separation.
The separation energy $(S_n)$, the Woods-Saxon potential parameters ($V_0$ and $V_\mathrm{so}$),
the excitation energy of $^{10}$Be $(\epsilon_2)$, the core-ground and core-excited probabilities $P_\mathrm{g.s.}(0)$ and $P_\mathrm{g.s.}(2)$,
and the exact and approximate BUXs ($\sigma$ and $\hat{\sigma}$) of $^{11}\mathrm{Be}+p$ at 63.7 MeV/nucleon are summarized.
All the values of $S_n$, $V_0$, $V_\mathrm{so}$, and $\epsilon_2$ are
shown in the units of MeV, whereas those of BUX are shown in millibarns.
Note that there are several resonances above $\varepsilon_\mathrm{th}=0.5$ MeV in set 7. See the text for details.
}
\label{tbl:summary}
\begin{center}
\begin{tabular}{ccccc c cccccc}
\hline\hline
 set & $S_n$ & $V_0$ & $V_\mathrm{so}$ & $\epsilon_2$ & $P_\mathrm{g.s.}(0)$ & $P_\mathrm{g.s.}(2)$ & $\sigma$(tot) & $\sigma(0)$ & $\hat{\sigma}(0)$ & $\sigma(2)$  & $\hat{\sigma}(2)$ \\
\hline
  1  & 0.1 & -51.924 & -8.5 & 3.368 & 0.943 & 0.057 & 92.4 & 82.8 & 79.4 &  9.6 & 13.0 \\
  2  & 0.5 & -54.45  & -8.5 & 3.368 & 0.855 & 0.145 & 54.8 & 47.8 & 44.3 &  7.0 & 10.5 \\
  3  & 0.5 & -52.988 & -1.0 & 0.5   & 0.792 & 0.208 & 56.0 & 45.8 & 44.6 & 10.2 & 11.4 \\
  4  & 1.0 & -56.475 & -8.5 & 3.368 & 0.788 & 0.212 & 48.8 & 42.6 & 39.6 &  6.2 &  9.2 \\
  5  & 5.0 & -67.059 & -8.5 & 3.368 & 0.577 & 0.423 & 10.4 &  7.6 &  6.5 &  2.8 &  3.9 \\
  6  & 5.0 & -65.670 & -1.0 & 0.5   & 0.545 & 0.455 &  7.6 &  4.0 &  3.4 &  3.6 &  4.2 \\
  7  & 1.0 & -85.791 & -1.0 & 0.5   & 0.679 & 0.321 & 24.2 & 13.7 & 12.4 & 10.5 & 11.8 \\
\hline\hline
\end{tabular}
\end{center}
\end{table*}

\begin{figure}[htbp]
\begin{center}
\includegraphics[width=0.45\textwidth,clip]{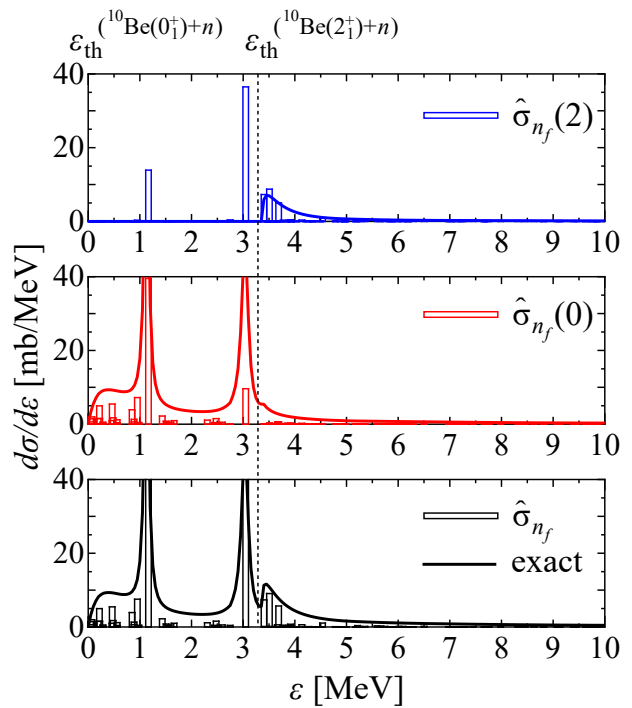}
\caption{Energy distribution of the BUXs for the scattering of $^{11}\mathrm{Be}+p$ at 63.7 MeV/nucleon.
The discretized and exact BUXs are represented by the bars and the line, respectively.
The total, core-ground, and core-excited BUXs are shown from the bottom panel.
Note that Eq.~\eqref{eq:imposition} is not imposed in this figure.
Each discretized BUX is multiplied by the density of states $\sum_{\ell_f j_f I_f}\bigl|\braket{\Psi_{JM}^{(\varepsilon_{n_f}\ell_f j_f I_f)}|\hat{\Psi}_{JM}^{(n_f)}}\bigr|^2$
 to make it comparable with the exact one in the same unit of millibarns/MeV.
}
\label{fig:discretizedBUXsmooth3}
\end{center}
\end{figure}

Here, we show that Eq.~\eqref{eq:imposition} plays a key role in the P separation.
In Fig.~\ref{fig:decomposition_both}, the open squares represent the results without Eq.~\eqref{eq:imposition}.
The results deviate from the exact ones significantly for $J^\pi=3/2^+$ and $5/2^+$.
The cause of this deviation is clearly seen in the energy distribution of the BUXs.
Figure~\ref{fig:discretizedBUXsmooth3} shows the discretized BUX together with the
exact BUX with reference to the $[{}^{10}\mathrm{Be}(0_1^+)+n]$-breakup threshold energy.
The $[{}^{10}\mathrm{Be}(2_1^+)+n]$-breakup threshold energy $(\varepsilon_\mathrm{th}=3.368$ MeV) is
also indicated by the vertical dotted line.
The bottom figure represents the total BUX $\hat{\sigma}_{n_f}$, and it is decomposed into 
the core-ground BUX $\hat{\sigma}_{n_f}(0)$ (middle) and the core-excited BUX $\hat{\sigma}_{n_f}(2)$ (top)
in accordance with Eq.~\eqref{eq:approx1} only, i.e., Eq.~\eqref{eq:imposition} is not imposed here.
Two peaks are seen at around 1.1~MeV ($5/2^+$) and 3.0~MeV ($3/2^+$),
and the corresponding states have the core-excited components as
 $P_\mathrm{res}(2)=0.293$ ($5/2^+$) and $P_\mathrm{res}(2)=0.794$ ($3/2^+$), respectively.
In particular, the $3/2^+$ state is the so-called Feshbach resonance
in which the core-excited component is dominant. However, those resonant states are never
broken up into ${}^{10}\mathrm{Be}(2_1^+)$ and $n$ because the core-excited channel is closed.
On the other hand, those resonant BUXs are counted as $\hat{\sigma}_{n_f}(2)$
by definition [Eq.~\eqref{eq:approx1}].
As a result, it turned out that Eq.~\eqref{eq:imposition} is important
in order to separate the total BUX properly.

{\it$^6$Li scattering in four-body CDCC.}
We then apply the P separation to $^6$Li scattering in the
framework of four-body CDCC ($n+p+\alpha+T$).
As discussed in the $^{11}$Be scattering,
we introduce the probability of the $d+\alpha$ configuration of $^6$Li~\cite{Wat15},
and define the $d\alpha$- and the $np\alpha$-BUXs
\eq{
\hat{\sigma}(d\alpha)&=\sum_{n_f} P_{n_f}(d\alpha) \hat{\sigma}_{n_f},\\
\hat{\sigma}(np\alpha)&=\sum_{n_f} \bigl[1-P_{n_f}(d\alpha)\bigr] \hat{\sigma}_{n_f},
}
respectively.
Following Eq.~\eqref{eq:imposition},
\be
P_{n_f}(d\alpha)=1\ \mathrm{and}\ P_{n_f}(np\alpha)=0
\ee
are imposed below the $n+p+\alpha$ threshold energy.
We predict the nuclear BUXs of ${}^6\mathrm{Li}+{}^{208}\mathrm{Pb}$ scattering at 39 and 210 MeV
for which the experimental data on elastic cross sections~\cite{Kee94,Nad89}
were well described by four-body CDCC~\cite{Wat15}.
The resultant BUXs are 
$\hat{\sigma}(d\alpha)=45.3$ and $\hat{\sigma}(np\alpha)=23.4$~mb for 39 MeV, and
$\hat{\sigma}(d\alpha)=89.9$ and $\hat{\sigma}(np\alpha)=47.1$~mb for 210 MeV.
The $\hat{\sigma}(np\alpha)$ constitutes one-third of the total BUX in the present energy range.
This appears to contradict with the findings in the previous work~\cite{Wat15}
that the four-body-channel-coupling effect is negligible in the elastic scattering.

In Ref.~\cite{Wat15}, the backcoupling effect on the elastic scattering was investigated
by switching on and off the three-body-channel coupling (${}^6\mathrm{Li}+T\leftrightarrow d+\alpha+T$)
and the four-body-channel coupling (${}^6\mathrm{Li}+T\leftrightarrow n+p+\alpha+T$).
Through the analysis, it was concluded that the three-body-channel coupling is essential
whereas the four-body-channel coupling is negligible in the elastic scattering.
Although this result seems to imply that $^6$Li is mostly broken up into $d+\alpha$
and hardly broken up into $n+p+\alpha$,
this has clearly been denied in the present results, i.e., $\hat{\sigma}(np\alpha)$ is
almost comparable with $\hat{\sigma}(d\alpha)$.
One of the possible interpretations of these results is that
$^6$Li may break up into three constituent particles
after breaking up into two clusters: ${}^6\mathrm{Li}\rightarrow d+\alpha \rightarrow n+p+\alpha$.

{\it Summary.}
We have proposed an approximate treatment (P separation) for decomposing
 discretized breakup cross sections into components of each channel.
We applied the P separation to $^{11}$Be scattering with core excitation
in which the core-ground and core-excited channels coexist.
The validity of the P separation is shown by demonstrating
that the approximate BUXs well reproduce the exact ones regardless of the configurations and/or the resonance positions of $^{11}$Be.
As a merit of the P separation, it is easily applied to four-body scattering
for which the exact breakup wave functions for the three-body projectile are difficult to obtain.
We have also applied the P separation to $^6$Li scattering in the $n+p+\alpha+T$ four-body model.
The predicted $np\alpha$-BUX is almost comparable with the $d\alpha$ BUX
contrary to the previous result in which the three-body-channel coupling was found to be essential
while the four-body-channel coupling is negligible. 
A possible way of understanding these results will be that $^6$Li breaks up into $n+p+\alpha$
mainly through the $(d+\alpha)$-breakup channel. We will investigate this hypothesis in a forthcoming paper.

{\it Acknowledgements.}
We would like to thank A.M. Moro for valuable discussion.
This work was supported by the Koshiyama Research Grant
and JSPS KAKENHI Grants No. JP16K05352 and No. JP18K03650.

%%--------------------------------------------------------------------%%
%%                           References                               %%
%%--------------------------------------------------------------------%%

\end{document}